\documentclass{kluwer}

\begin{document}      

\begin{article}                                                     
\begin{opening}         
\title{TILTED BIANCHI TYPE I  COSMOLOGICAL MODELS FILLED WITH 
DISORDERED RADIATION IN GENERAL RELATIVITY REVISITED }
\author{ANIRUDH PRADHAN\thanks{Corresponding Author}}
\runningtitle{TILTED BIANCHI TYPE I  COSMOLOGICAL MODELS}
\runningauthor{A. PPRADHAN AND ABHA RAI}
\author{ABHA RAI}
\institute{Department of Mathematics, Hindu Post-graduate College,
Zamania-232 331, Ghazipur, U. P., India; \\
E-mail:acpradhan@yahoo.com, apradhan@mri.ernet.in, pradhan@iucaa.ernet.in}

\date{\today}

\begin{abstract} 
Tilted Bianchi type I cosmological models filled with disordered radiation
in presence of a bulk viscous fluid and heat flow are investigated. The 
coefficient of bulk viscosity is assumed to be a power function of mass density. 
Some physical and geometric properties of the models are also discussed.
\end{abstract}
\keywords{ Cosmology; Bianchi type I Universe; Tilted  Models.\\ }

\end{opening}

\section{Introduction}
\vspace*{-0.5pt}
\noindent
General Relativity describes the state in which radiation concentrates
around a star. Klein (1948) worked on it and obtained an approximate 
solution to Einsteinian field equations in spherical symmetry for a 
distribution of diffused radiation. Many other researchers (Singh and 
Abdussattar, 1973; Roy and Bali, 1977) have worked on this topic and  
obtained exact static spherically and cylindrically symmetric solutions 
of Einstein's field equations with exception as well. Roy and Singh (1977) 
have obtained a non-static plane symmetric spacetime filled with disordered 
radiation. Teixeira, Wolk and Som (1977) investigated a model filled with
source free disordered distribution of electromagnetic radiation in 
Einstein's general relativity.\\

The general dynamics of tilted models have been studied by King and Ellis 
(1973) and Ellis and King (1974). The cosmological models with heat flow 
have been also studied by Coley and Tupper (1983. 1984); Roy and Banerjee 
(1988). Ellis and Baldwin (1984) have shown that we are likely to be living 
in a tilted universe and they have indicated how we may detect it. Beesham
(1986) derived tilted Bianchi type V cosmological models in the scale-covariant
theory. A tilted cold dark matter cosmological scenario has been discussed
by Cen, Nickolay, Kofman and Ostriker (1992).\\

The majority of the studies in cosmology involve a perfect fluid. However, 
observed physical phenomena such as the large entropy per baryon and the 
remarkable degree of isotropy of the cosmic microwave background radiation 
suggests analysis of dissipative effects in cosmology. Furthermore, there are
several processes which are expected to give rise to viscous effects. These
are the decoupling of neutrinos during the radiation era and the decoupling
of radiation and matter during the recombination era. Bulk viscosity is
associated with the GUT phase transition and string creation. The model 
studied by Murphy (1973) possessed an interesting feature in that the 
big bang type of singularity of infinite spacetime curvature does not occur 
to be a finite past. However, the relationship assumed by Murphy between the
viscosity coefficient and the matter density is not acceptable at large 
density. The effect of bulk viscosity on the cosmological evolution has been 
investigated by a number of authors in the framework of general theory of 
relativity (Padmanabhan and Chitre, 1987; Johri and Sudarshan, 1988; Maartens, 
1995; Zimdahl, 1996; Pradhan, Sarayakar and Beesham, 1997; Kalyani and Singh 
(1997; Singh, Beesham and Mbokazi, 1998; Pradhan et al., 2001, 2002). This 
motivates to study cosmological bulk viscous fluid model.\\

Recently Bali and Meena (2002) have investigated two tilted cosmological models 
filled with disordered radiation of perfect fluid and heat flow. Meena and Bali
(2002) have obtained two conformally flat tilted Bianchi type V cosmological 
models. Very recently tilted Bianchi type I cosmological model for perfect
fluid distribution in presence of magnetic field is investigated by Bali and
Sharma (2003). In this paper, we propose to find tilted Bianchi type I cosmological 
models filled with disordered radiation in presence of a bulk viscous fluid and 
heat flow.\\   
\section{Field Equations}
\noindent
We consider the Bianchi type I metric in the form
\begin{equation}
\label{eq1}
ds^{2} = - dt^{2} + A^{2} dx^{2} + B^{2}dy^{2} + C^{2}dz^{2},
\end{equation}
where A, B and C are function of $t$ only.\\
The Einstein's field equations are given by
\begin{equation}
\label{eq2}
R^{j}_{i} - \frac{1}{2} R g^{j}_{i} = -8\pi T^{j}_{i}~ ~ 
\mbox{(c = 1, G = 1 in gravitational unit)}
\end{equation}
where $R^{j}_{i}$ is the Ricci tensor; $R$ = $g^{ij} R_{ij}$ is the
Ricci scalar; and $T^{j}_{i}$ is the stress energy-tensor in the presence
of bulk stress given by 
\begin{equation}
\label{eq3}
T^{j}_{i} = (\rho + \bar{p})v_{i}v^{j} + \bar{p} g^{j}_{i} + 
q_{i}v^{j} + v_{i}q^{j},
\end{equation}
and
\begin{equation}
\label{eq4}
\bar{p} = p - \xi v^{i}_{;i}.
\end{equation}
Here $\rho$, $p$, $\bar{p}$ and $\xi$ are the energy density,
isotropic pressure, effective pressure, flow vector, bulk viscous 
coefficient respectively and $v_{i}$ is the flow vector satisfying 
the relations
\begin{equation}
\label{eq5}
g_{ij} v^{i}v^{j} = - 1
\end{equation}
\begin{equation}
\label{eq6}
q_{i} q^{j} > 0,
\end{equation}
\begin{equation}
\label{eq7}
q_{i}v^{i} = 0,
\end{equation}
where $q_{i}$ is the  heat conduction vector orthogonal to $v_{i}$.
The fluid flow vector has the component $(\frac{\sinh  \lambda}{A}, 0, 0, 
\cosh  \lambda)$ satisfying Eq. (\ref{eq5}) and $\lambda$ is the tilt angle.\\
The Einstein's field equations (\ref{eq2}) for the line element (\ref{eq1})
has been set up as
\begin{equation}
\label{eq8}
- 8\pi[(\rho + \bar{p})\sinh^{2}  \lambda + \bar{p} + 2q_{1}\frac{\sinh  \lambda}
{A}] = \frac{B_{44}}{B} + \frac{C_{44}}{C} + \frac{B_{4}C_{4}}{BC}, 
\end{equation}
\begin{equation}
\label{eq9}
-8\pi \bar{p} = \frac{A_{44}}{A} + \frac{C_{44}}{C} + \frac{A_{4}C_{4}}{AC}, 
\end{equation}
\begin{equation}
\label{eq10}
-8\pi \bar{p} = \frac{A_{44}}{A} + \frac{B_{44}}{B} + \frac{A_{4}B_{4}}{AB}, 
\end{equation}
\begin{equation}
\label{eq11}
- 8\pi[- (\rho + \bar{p})\cosh^{2}  \lambda + \bar{p} - 2q_{1}\frac{\sinh  \lambda}
{A}] = \frac{A_{4}B_{4}}{AB} + \frac{A_{4}C_{4}}{AC} + \frac{B_{4}C_{4}}{BC}, 
\end{equation}
\begin{equation}
\label{eq12}
(\rho + \bar{p})A \sinh  \lambda ~  \cosh  \lambda + q_{1}\cosh  \lambda +
q_{1}\frac{\sinh^{2}  \lambda}{\cosh  \lambda} = 0,
\end{equation}
where the suffix $4$ at the symbols $A$, $B$, $C$ denotes ordinary 
differentiation with respect ti $t$. 
\section{Solutions of the field equations}
Equations (\ref{eq8}) - (\ref{eq12}) with Eq. (\ref{eq4}) are five independent 
equations in eight unknowns $A$, $B$, $C$, $\rho$, $p$, $\xi$, $q$ and $\lambda$.
For the complete determinacy of the system, we need three extra conditions.\\
We assume the model is filled with disordered radiation which leads to
\begin{equation}
\label{eq13}
\rho = 3 p
\end{equation}
and 
\begin{equation}
\label{eq14}
A = (B ~ C)^{n}
\end{equation}
where $n$ is any positive real number.
Eqs. (\ref{eq8}) and (\ref{eq11}) lead to
\begin{equation}
\label{eq15}
\frac{B_{44}}{B} + \frac{C_{44}}{C} + \frac{2B_{4} C_{4}}{BC} +
\frac{A_{4}C_{4}}{AC} + \frac{A_{4}B_{4}}{AB} = 8\pi (\rho - p + 2\xi \theta)
\end{equation}
Using Eq. (\ref{eq13}) in Eq. (\ref{eq15}) reduces to
\begin{equation}
\label{eq16}
\frac{B_{44}}{B} + \frac{C_{44}}{C} + \frac{2B_{4} C_{4}}{BC} +
\frac{A_{4}C_{4}}{AC} + \frac{A_{4}B_{4}}{AB} = 16 \pi ( p - \xi \theta)
\end{equation}
Equations (\ref{eq9}) and (\ref{eq10}) lead to
\begin{equation}
\label{eq17}
\frac{C_{44}}{C} - \frac{B_{44}}{B} + \frac{A_{4}}{A}(\frac{C_{4}}{C} 
- \frac{B_{4}}{B}) = 0
\end{equation}
which leads to
\begin{equation}
\label{eq18}
\frac{\mu_{4}}{\mu} = \frac{k}{\epsilon^{n + 1}}
\end{equation}
where $BC = \epsilon$, $\frac{B}{C} = \nu$ and $k$ is a constant of integration.\\
Eqs. (\ref{eq9}) and (\ref{eq10}) also give
\begin{equation}
\label{eq19}
\frac{2A_{44}}{A} + \frac{B_{44}}{B} + \frac{C_{44}}{C} +
\frac{A_{4}C_{4}}{AC} + \frac{A_{4}B_{4}}{AB} = - 16 \pi (p - 2\xi \theta)
\end{equation}
From Eqs. (\ref{eq16}) and (\ref{eq19}), we obtain 
\begin{equation}
\label{eq20}
\frac{A_{44}}{A} + \frac{B_{44}}{B} + \frac{C_{44}}{C} +
+ \frac{B_{4}C_{4}}{BC} + \frac{A_{4}C_{4}}{AC} + \frac{A_{4}B_{4}}{AB} = 0
\end{equation}
which can be rewritten as
\begin{equation}
\label{eq21}
(n + 1)\frac{\epsilon_{44}}{\epsilon} + n^{2} \frac{\epsilon^{2}_{4}}
{\epsilon^{2}} - \frac{1}{4}\frac{\epsilon^{2}_{4}}{\epsilon^{2}} 
+ \frac{1}{4}\frac{\nu^{2}_{4}}{\nu^{2}} = 0
\end{equation}
where $A = \epsilon^{n}$. From Eqs. (\ref{eq18}) and (\ref{eq21}), we get 
\begin{equation}
\label{eq22}
f f^{1} + \frac{(4n^{2} - 1)f^{2}}{4(n + 1)\epsilon} = \frac{k^{2}}{4(n + 1)
\epsilon^{2n + 1}}
\end{equation}
where $\epsilon_{4} = f(\epsilon)$. Eq. (\ref{eq22}) leads to
\begin{equation}
\label{eq23}
f^{2} = \frac{k^{2} + (4n + 1)k_{1}\epsilon^{\frac{4n + 1}{2(n + 1)}}}
{(4n + 1)\epsilon^{2n}}
\end{equation}
where $k_{1}$ ia a constant of integration. On integrating Eq. (\ref{eq18})
we obtain
\begin{equation}
\label{eq24}
log~\nu = k\sqrt{(4n + 1)} \int{\frac{d\epsilon}{\epsilon \sqrt{k^{2} + k_{1}
(4n + 1)\epsilon^{\frac{4n + 1}{2(n + 1)}}}}}
\end{equation}
Hence the metric (\ref{eq1}) reduces to the form
\begin{equation}
\label{eq25}
ds^{2} = -\frac{d\epsilon^{2}}{f^{2}} + \epsilon^{2n} dx^{2} + 
\epsilon~ \nu~ dy^{2} + \frac{\epsilon}{\nu}dz^{2} 
\end{equation}
where $\nu$ is determined by Eq. (\ref{eq24}).\\
By using the following transformation
\[
\epsilon = T, ~ ~ 
x = X, ~ ~ 
y = Y, ~ ~ 
z = Z \]
the metric (\ref{eq25}) takes the form
\begin{equation}
\label{eq26}
ds^{2} = - \left[\frac{(4n + 1)T^{2n}}{k^{2} + (4n + 1)k_{1}T^{\frac{4n + 1}
{2(n + 1)}}}\right] dT^{2} + T^{2n}~ dX^{2} + T\nu~ dY^{2} + 
\frac{T}{\nu}~ dZ^{2}
\end{equation}
where
\begin{equation}
\label{eq27}
log~\nu = k\sqrt{(4n + 1)} \int{\frac{dT}{T \sqrt{k^{2} + k_{1}
(4n + 1)T^{\frac{4n + 1}{2(n + 1)}}}}}
\end{equation} 
The pressure and density of the model (\ref{eq26}) are given by
\begin{equation}
\label{eq28}
8\pi (p -\xi \theta) = \frac{(4n + 1)k_{1}}{8(n + 1)T^{\frac{4n^{2} + 4n +3}
{2(n + 1)}}}
\end{equation} 
\begin{equation}
\label{eq29}
8\pi (\rho - 3\xi \theta) = \frac{3(4n + 1)k_{1}}{8(n + 1)
T^{\frac{4n^{2} + 4n +3}{2(n + 1)}}}
\end{equation} 
The tilt angle $\lambda$ is given by
\begin{equation}
\label{eq30}
\cosh^{2}  \lambda = \frac{2n + 3}{2(2n + 1)}
\end{equation} 
Hence
\begin{equation}
\label{eq31}
\sinh^{2}  \lambda = \frac{1 - 2n}{2(2n + 1)}
\end{equation} 
The scalar of the expansion $\theta$, calculated for the flow vector 
$v^{i}$, is given by
\begin{equation}
\label{eq32}
\theta = \frac{(n + 1)}{T^{n + 1}}\sqrt{\frac{(2n + 3)[k^{2} + (4n + 1)k_{1}
T^{\frac{4n + 1}{2(n + 1)}}]}{2(2n + 1)(4n + 1)}}
\end{equation} 
If we put $\xi = 0$ in Eqs. (\ref{eq28})-(\ref{eq29}), we get the solutions
as obtained by Bali and Meena (2002).\\
Thus, given $\xi(t)$ we can solve the system for the physical quantities.
Therefore to apply the third condition, let us assume the following {\it adhoc}
law (Maartens, 1995; Zimdahl, 1996) 
\begin{equation}
\label{eq33}
\xi(t) = \xi_{0} \rho^{m}
\end{equation} 
where $\xi_{0}$ and $m$ are real constants. If $m = 1$, Eq. (\ref{eq33})
may correspond to a radiative fluid (Weinberg, 1972), whereas 
$m$ = $\frac{3}{2}$ may correspond to a string-dominated universe. However, 
more realistic models (Santos, 1985) are based on lying the regime 
$0 \leq m \leq \frac{1}{2}$.    
\subsection {Model I: ~ ~ ~ $(\xi = \xi_{0})$}
When $m = 0$, Eq. (\ref{eq33}) reduces to $\xi$ = $\xi_{0}$ and hence 
Eqs. (\ref{eq28}) and (\ref{eq29}) with the use of Eq. (\ref{eq32}) lead to
\begin{equation}
\label{eq34}
p = \frac{(n + 1)\xi_{0}}{T^{n + 1}}\sqrt{\frac{(2n + 3)[k^{2} + (4n + 1)k_{1}
T^{\frac{4n + 1}{2(n + 1)}}]}{2(2n + 1)(4n + 1)}} + \frac{(4n + 1)k_{1}}{64(n + 1)
\pi T^{\frac{4n^{2} + 4n +3}{2(n + 1)}}}
\end{equation} 
\begin{equation}
\label{eq35}
\rho = \frac{(n + 1)\xi_{0}}{3T^{n + 1}}\sqrt{\frac{(2n + 3)[k^{2} + (4n + 1)k_{1}
T^{\frac{4n + 1}{2(n + 1)}}]}{2(2n + 1)(4n + 1)}} + \frac{(4n + 1)k_{1}}{192(n + 1)
\pi T^{\frac{4n^{2} + 4n +3}{2(n + 1)}}}
\end{equation} 
\subsection {Model II: ~ ~ ~ $(\xi = \xi_{0}\rho)$}
When $m = 0$, Eq. (\ref{eq33}) reduces to $\xi$ = $\xi_{0}\rho$ and hence 
Eqs. (\ref{eq28}) and (\ref{eq29}) with the use of Eq. (\ref{eq32}) lead to
\begin{equation}
\label{eq36}
p = \frac{(4n + 1)k_{1}}{64 \pi (n + 1)T^{\frac{2n^{2} + 1}{2(n + 1)}}
\left[T^{(n + 1)} - 3(n + 1)\xi_{0}\sqrt{(2n + 3)\left[k^{2} + 
(4n + 1)k_{1}T^{\frac{4n + 1}{2(n + 1)}}\right]}\right]}
\end{equation} 
\begin{equation}
\label{eq37}
\rho = \frac{(4n + 1)k_{1}}{192 \pi (n + 1)T^{\frac{2n^{2} + 1}{2(n + 1)}}
\left[T^{(n + 1)} - 3(n + 1)\xi_{0}\sqrt{(2n + 3)\left[k^{2} + 
(4n + 1)k_{1}T^{\frac{4n + 1}{2(n + 1)}}\right]}\right]}
\end{equation} 
\section{Some Physical and Geometric Properties of the Models}
The flow vector $v^{i}$ and heat conduction vector $q^{i}$ for the models 
(\ref{eq26}) are obtained by Bali and Meena (2002)
\begin{equation}
\label{eq38}
v^{1} = \frac {1}{T^{n}}\sqrt{\frac{(1 - 2n)}{2(2n + 1)}},
\end{equation} 
\begin{equation}
\label{eq39}
v^{4} = \sqrt{\frac{2n + 3}{2(2n + 1)}},
\end{equation} 
\begin{equation}
\label{eq40}
q^{1} = - \frac {b(4n + 1)(2n + 3)}{64 \pi (n + 1)T^{\frac{4n^{2} - 4n +3}
{2(n + 1)}}} \sqrt{\frac{(1 - 2n)}{2(2n + 1)}},
\end{equation} 
\begin{equation}
\label{eq41}
q^{4} = \frac {b(4n + 1)^{2} (1 - 2n)}{64 \pi (n + 1)}\sqrt{\frac{2n + 3}
{2(2n + 1)}}\left[\frac{T^{\frac{2n^{2} + 2n -3}{2(n + 1)}}}{a^{2} + b(4n + 1)
T^{\frac{4n + 1}{2(n + 1)}}}\right].
\end{equation} 
The non-vanishing components of shear tensor $(\sigma_{ij})$ and rotation
tensor $(\omega_{ij})$ are obtained as
\begin{equation}
\label{eq42}
\sigma_{11} = \frac{(2n - 1)}{6}\left[\frac{2n + 3}{2n + 1}\right]^{\frac{3}{2}}
T^{n - 1}\sqrt {\frac{a^{2} + b(4n + 1)T^{\frac{4n + 1}{2(n + 1)}}}
{2(4n + 1)}},
\end{equation} 
\begin{equation}
\label{eq43}
\sigma_{22} = \nu \sqrt{\frac{2n + 3}{2(2n + 1)(4n + 1)}}\left[ \frac{3a\sqrt{4n + 1}
+ (1 - 2n)\sqrt{a^{2} + b(4n + 1) T^{\frac{4n + 1}{2(n + 1)}}}}{6T^{n}}\right],
\end{equation} 
where $\nu$ is already given by (\ref{eq27}).
\begin{equation}
\label{eq44}
\sigma_{33} = \frac{1}{\nu} \sqrt{\frac{2n + 3}{2(2n + 1)(4n + 1)}}
\left[ \frac{(1 - 2n)\sqrt{a^{2} + b(4n + 1) 
T^{\frac{4n + 1}{2(n + 1)}}} - 3a\sqrt{4n + 1}}{6T^{n}}\right],
\end{equation} 
\begin{equation}
\label{eq45}
\sigma_{44} = - \frac{(1 - 2n)^{2}}{6(2n + 1)} \sqrt{\frac{2n + 3}
{2(2n + 1)(4n + 1)}} \left[ \frac{\sqrt{a^{2} + b(4n + 1) 
T^{\frac{4n + 1}{2(n + 1)}}}}{T^{n + 1}}\right],
\end{equation} 
\begin{equation}
\label{eq46}
\sigma_{14} = \frac{(6 - 2n -32n^{2})}{12(2n + 1)^{\frac{3}{2}}} 
\sqrt{\frac{(1 - 2n)}{2(4n + 1)}}
\left[ \frac{\sqrt{a^{2} + b(4n + 1) 
T^{\frac{4n + 1}{2(n + 1)}}}}{T}\right],
\end{equation} 
\begin{equation}
\label{eq47}
\omega_{14} = n \sqrt{\frac{(1 - 2n)}{2(2n + 1)(4n + 1)}}
\left[ \frac{\sqrt{a^{2} + b(4n + 1) 
T^{\frac{4n + 1}{2(n + 1)}}}}{T}\right].
\end{equation} 
The rates of expansion $H_{i}$ in the direction of $X$, $Y$, $Z$-axes
are given by
\begin{equation}
\label{eq48}
H_{1} = \frac{n}{\sqrt{4n + 1}}
\left[ \frac{\sqrt{a^{2} + b(4n + 1) 
T^{\frac{4n + 1}{2(n + 1)}}}}{T^{n + 1}}\right].
\end{equation} 
\begin{equation}
\label{eq49}
H_{2} = \left[ \frac{\sqrt{a^{2} + b(4n + 1) 
T^{\frac{4n + 1}{2(n + 1)}}} + a \sqrt{4n + 1}}{2\sqrt{4n + 1}T^{n + 1}}\right],
\end{equation} 
\begin{equation}
\label{eq50}
H_{2} = \left[ \frac{\sqrt{a^{2} + b(4n + 1) 
T^{\frac{4n + 1}{2(n + 1)}}} - a \sqrt{4n + 1}}{2\sqrt{4n + 1}T^{n + 1}}\right],
\end{equation} 
The models in general represent shearing and rotating universes. The expansion 
in the models decreases as time increases and the expansion in the models stops 
at $T = \infty$. There is a big bang in the models at $T= 0$ if $n + 1
> 0$. There is no rotation in the models for $n = 0$ but it is shearing and goes
on decreasing as time increases. For $n = 0$, the Hubble constant $H_{1}$ = $0$,
$H_{2} \ne 0$, $H_{3} \ne 0$. The fluid velocity vectors $v^{1}$ and $v^{4}$
for $n$ = $0$, are given by $v^{1}$ = $\frac{1}{\sqrt{2}}$, $v^{4}$ = 
$\sqrt{\frac{3}{2}}$.  Both density and pressure in the models become zero at 
$ T = \infty$. Since $\lim_{t\rightarrow \infty}
\frac{\sigma}{\theta} \ne 0$, the models do not approach isotropy for large values 
of $T$ for general value of $n $ and $n$ = $0$ also.
\section{Particular Models }
We consider the metric
\begin{equation}
\label{eq51}
ds^{2} = - dt^{2} + dx^{2} + B^{2} dy^{2} + C^{2}dz^{2},
\end{equation} 
where $B$, $C$ are functions of $t$ alone.\\
Here we only assume that the model is filled with disordered radiation which 
leads Eq. (\ref{eq13}). Following Bali and Meena (2002), we obtain, from Eqs.
(\ref{eq51}), (\ref{eq2}) and (\ref{eq3})
\begin{equation}
\label{eq52}
B^{2} = \frac{k_{1} (f - k)^{4}}{k^{2}_{2}},
\end{equation} 
\begin{equation}
\label{eq53}
C^{2} = \frac{f + k)}{k_{1} k^{2}_{2}},
\end{equation} 
where $k$, $k_{1}$ and $k_{2}$ are constants of integration. Here $BC$ = 
$\epsilon$, $\frac{B}{C}$ = $\nu$, $\epsilon_{4}$ = $f(\epsilon)$ and
\begin{equation}
\label{eq54}
\epsilon = (\frac{f^{2} k^{2}}{k_{2}})^{2}
\end{equation} 
\begin{equation}
\label{eq55}
\nu = k_{1}(\frac{f - k}{f + k})^{2}
\end{equation} 
Thus, after the suitable transformation of coordinates, the metric (\ref{eq51})
reduces to the form
\begin{equation}
\label{eq56}
ds^{2} = \frac{16}{k^{4}_{2}}(T^{2} - k^{2})^{2} dT^{2} + dX^{2} + (T - k)^{2}dY^{2}
+ (T + k)^{4} dZ^{2}
\end{equation} 
The pressure and density for the model (\ref{eq56}) are given by
\begin{equation}
\label{eq57}
\bar{p} = (p - \xi \theta) = \frac{k^{4}_{2}}{64 \pi (T^{2} - k^{2})^{3}}
\end{equation} 
\begin{equation}
\label{eq58}
(\rho - 3 \xi \theta) = \frac{3 k^{4}_{2}}{64 \pi (T^{2} - k^{2})^{3}}
\end{equation} 
The tilt angle $\lambda$ is given by
\begin{equation}
\label{eq59}
\cosh  \lambda = \sqrt{\frac{3}{2}} 
\end{equation} 
The scalar expansion $\theta$ calculated for the flow vector $v^{i}$,
is given by
\begin{equation}
\label{eq60}
\theta = \sqrt{\frac{3}{2}}\left[\frac{k^{2}_{2}T}{(T^{2} - k^{2})^{3}}\right]
\end{equation} 
Thus, given $\xi(t)$ one can solve the system for the physical quantities.
\subsection{Model I: $ (\xi = \xi_{0})$ }
When $m$ = $0$, Eq. (\ref{eq33}) reduces to $\xi$ = $\xi_{0}$ and hence
Eqs. (\ref{eq57}) and (\ref{eq58}) woth the use of Eq. (\ref{eq60}) lead to
\begin{equation}
\label{eq61}
p = \frac{k^{2}_{2}}{(T^{2} - k^{2})^{2}}\left[\sqrt{\frac{3}{2}}~\xi_{0} T +
\frac{k^{2}_{2}}{64 \pi (T^{2} - k^{2})}\right]
\end{equation} 
\begin{equation}
\label{eq62}
\rho = \frac{3 k^{2}_{2}}{(T^{2} - k^{2})^{2}}\left[\sqrt{\frac{3}{2}}~\xi_{0} T +
\frac{k^{2}_{2}}{64 \pi (T^{2} - k^{2})}\right]
\end{equation} 
\subsection{Model I: $ (\xi = \xi_{0}\rho)$ }
When $m$ = $1$, Eq. (\ref{eq33}) reduces to $\xi$ = $\xi_{0}\rho$ and hence 
Eqs. (\ref{eq57}) and (\ref{eq58}) with the help of Eq. (\ref{eq60}) lead to
\begin{equation}
\label{eq63}
p = \frac{k^{4}_{2}}{32 \pi (T^{2} - k^{2})\left[2(T^{2} - k^{2})^{2} - 3\sqrt{6}~
k^{2}T\right]}
\end{equation} 
\begin{equation}
\label{eq64}
p = \frac{3 k^{4}_{2}}{32 \pi (T^{2} - k^{2})\left[2(T^{2} - k^{2})^{2} - 3\sqrt{6}~
k^{2}T\right]}
\end{equation} 
\section{Some Physical and Geometric Properties of Particular Models}
The non-vanishing components of shear tensor $\sigma-{ij}$ and rotation tensor
$\omega_{ij}$ for the models (\ref{eq56}) are given by (Bali and Meena, 2002)
\begin{equation}
\label{eq65}
\sigma_{11} = - \frac{1}{2}\sqrt{\frac{3}{2}}\left[\frac{k_{2}^{2} T}{(T^{2} 
- k^{2})^{2}}\right],
\end{equation} 
\begin{equation}
\label{eq66}
\sigma_{22} =  \frac{k_{1}}{6}\sqrt{\frac{3}{2}}\left[\frac{(T + 3k)(T - k)^{2}}
{(T + k)^{2}}\right]
\end{equation} 
\begin{equation}
\label{eq67}
\sigma_{33} =  \frac{1}{6 k}\sqrt{\frac{3}{2}}\left[\frac{(T - 3k)(T + k)^{2}}
{(T - k)^{2}}\right]
\end{equation} 
\begin{equation}
\label{eq68}
\sigma_{44} = - \frac{1}{6}\sqrt{\frac{3}{2}}\left[\frac{k_{2}^{2} T}{(T^{2} 
- k^{2})^{2}}\right],
\end{equation} 
\begin{equation}
\label{eq69}
\sigma_{14} = \frac{1}{2\sqrt{2}}\left[\frac{k_{2}^{2} T}{(T^{2} 
- k^{2})^{2}}\right],
\end{equation} 
\begin{equation}
\label{eq70}
\omega_{14} = 0
\end{equation} 
The expressions for $v_{1}$, $v_{4}$, $q_{1}$ and $q_{4}$ for the metric 
(\ref{eq56}) are obtained as
\begin{equation}
\label{eq71}
v_{1} = \frac{1}{\sqrt{2}},
\end{equation} 
\begin{equation}
\label{eq72}
v_{4} = - \frac{16}{k_{2}^{4}}(T^{2} - k^{2})^{2}\sqrt{\frac{3}{2}},
\end{equation} 
\begin{equation}
\label{eq73}
q_{1} = - \frac{3k_{2}^{4}}{64 \sqrt{2}~ \pi(T^{2} - a^{2})^{3}},
\end{equation} 
\begin{equation}
\label{eq74}
q_{4} =  \frac{3k_{2}^{4}}{64 \sqrt{6}~ \pi(T^{2} - a^{2})^{3}}.
\end{equation} 
The rate of expansion $H_{i}$ in the directions of $X$, $Y$, $Z$-axes is given by
\begin{equation}
\label{eq75}
H_{1} = 0,
\end{equation} 
\begin{equation}
\label{eq76}
H_{2} = \frac{k_{2}^{2}}{2(T + k) (T - k)^{2}},
\end{equation} 
\begin{equation}
\label{eq77}
H_{3} = \frac{k_{2}^{2}}{2(T - k) (T + k)^{2}}.
\end{equation} 
The models start with a big bang at $T$ = $ \pm k$ and the expansion in the models
decreases as time increases and the expansion in the models stops at $T$ = $0$.
The x-component of the Hubble parameter is zero due to the assumption of metric.
However, the $y$ and $x$ components become infinite at $T$ = $k$. The models, 
in general, represent shearing and non-rotating universe.\\
Since \\  
\[
 \lim_{t\rightarrow \infty} \frac{\sigma}{\theta} \ne 0. 
\]
Hence the models do not approach isotropy for large values of $T$. When 
$T \rightarrow k$, $q_{1}\rightarrow -\infty$ and $q_{4}$ is constant. There is 
a singularity in the models at $T$ = $k$. This singularity is pan cake type 
(MacCallum, 1971) as $g_{11}$ and $g_{22}$ is zero at the singularity $T$ = $k$. 
\section*{Acknowledgements} 
\noindent
A. Pradhan thanks to the Inter-University Centre for Astronomy and Astrophysics, 
India for providing  facility under Associateship Programmes where this work was 
carried out. Authors would also like to thank Raj Bali for helpful discussions.  \\
\newline
\newline

\end{article}
\end{document}